\documentclass[twocolumn,prb,amsmath,amssymb,floatfix]{revtex4-2}
\usepackage{amsmath}
\usepackage{amssymb}
\usepackage{graphicx}
\usepackage{bm}
\usepackage{color}
\usepackage{hyperref}
\usepackage{graphicx}
\usepackage{upgreek}
\usepackage{scalerel}
\usepackage{multirow}
\usepackage{pgffor}
\usepackage{pdfpages}
\usepackage{mathdots}
\usepackage{float}
\usepackage{silence}
\usepackage{physics} 
\usepackage{lscape}   
\usepackage{color, colortbl}
\usepackage[table]{xcolor}
\usepackage{comment}  
\makeatletter
\AtBeginDocument{\let\LS@rot\@undefined}
\makeatother

\begin{document}

\title{Bulk Bogoliubov Fermi arcs in non-Hermitian superconducting systems}
\author{Jorge Cayao}
%\email[]{jorge.cayao@physics.uu.se}
\affiliation{Department of Physics and Astronomy, Uppsala University, Box 516, S-751 20 Uppsala, Sweden}
\author{Annica M. Black-Schaffer}
\affiliation{Department of Physics and Astronomy, Uppsala University, Box 516, S-751 20 Uppsala, Sweden}

\date{\today}
\begin{abstract}
We consider a non-Hermitian superconducting system by coupling a conventional superconductor to a ferromagnet lead and demonstrate the emergence of exceptional points when an external Zeeman field is applied. We  discover that, depending on the non-Hermiticity and the Zeeman field, the exceptional points mark the ends of lines with zero real energy, thus giving rise to topologically protected and highly tunable bulk Fermi arcs, which we coin bulk Bogoliubov Fermi arcs due to their superconducting nature.  We show that these bulk Bogoliubov Fermi arcs are the non-Hermitian counterparts of the Hermitian topological phase transition but are much more prevalent and also experimentally detectable through large spectral signatures. 
\end{abstract}

\maketitle
%%%%%%%%%%%%%%%%%%%%%%%%%%%%%%%
% SECTION 1:                 INTRODUCTION                                %
%%%%%%%%%%%%%%%%%%%%%%%%%%%%%%%

\section{Introduction}
\label{section:Intro}
Topological materials have spurred large interest in the last decade, not only because they represent new states of matter but also due to their potential applications \cite{doi:10.7566/JPSJ.82.102001, seidel2019nanoelectronics, liu2019topological, gilbert2021topological}. While initial studies focused on isolated systems described by Hermitian models \cite{RevModPhys.83.1057,sato2017topological,RevModPhys.88.035005}, topological concepts have more recently been extended to open and dissipative setups modeled by effective non-Hermitian (NH) Hamiltonians \cite{PhysRevX.8.031079, PhysRevB.99.235112, PhysRevX.9.041015}. 
Interestingly, NH systems have been shown to realize unexpected topological phases in systems ranging from classical metamaterials to condensed matter setups with no Hermitian analogs \cite{el2018non,ozdemir2019parity,RevModPhys.93.015005,doi:10.1080/00018732.2021.1876991,wiersig2020review, parto2020non}.

The intriguing topological properties of NH systems occur due to a class of degeneracies known as exceptional points (EPs) \cite{TKato, heiss2004exceptional, berry2004physics, Heiss_2012, PhysRevLett.86.787, PhysRevLett.103.134101, PhysRevLett.104.153601, gao2015observation, doppler2016dynamically,PhysRevB.99.121101}, where both eigenvalues and eigenvectors coalesce. The emergent NH topology due to EPs has already been shown to lead to a plethora of topological phenomena \cite{RevModPhys.93.015005,doi:10.1080/00018732.2021.1876991}, such as enhanced sensing performance \cite{hodaei2017enhanced,chen2017exceptional}, unidirectional lasing \cite{peng2016chiral,Longhi:17}, and bulk Fermi arcs \cite{kozii2017non,PhysRevB.97.014512,PhysRevB.98.035141,PhysRevB.99.041202,PhysRevLett.123.066405,PhysRevLett.123.123601,science359Zhou,doi:10.7566/JPSCP.30.011098, PhysRevLett.125.227204, PhysRevLett.127.186601,PhysRevLett.127.186602}, none with a Hermitian counterpart. The bulk Fermi arcs are particularly fascinating as they are genuine bulk topological states that appear at zero real energy connecting EPs in momentum space, unlike
  the standard topological zero-energy states in Hermitian systems only emerging at boundaries \cite{RevModPhys.83.1057,sato2017topological,RevModPhys.88.035005}.  
 
 Bulk Fermi arcs have so far been primarily studied in normal-state systems \cite{kozii2017non,PhysRevB.98.035141,PhysRevLett.123.066405,PhysRevLett.123.123601,science359Zhou,doi:10.7566/JPSCP.30.011098, PhysRevLett.125.227204, PhysRevLett.127.186601,PhysRevLett.127.186602}, while only limited evidence exists in superconductors, and then often using rather unrealistic sources of non-Hermiticity, such as an unconventional order parameter \cite{PhysRevB.97.014512,PhysRevB.99.041202} or a need for Dirac materials \cite{PhysRevB.99.165145,PhysRevB.103.235438}. A more realistic and experimentally relevant platform for NH physics is simple material junctions between conventional superconductors and normal metals \cite{datta1997electronic,PhysRevB.98.155430, PhysRevB.98.245130, PhysRevResearch.1.012003}, which  have already proven powerful for revealing boundary states \cite{pikulin2012topological, PhysRevB.87.235421,ioselevich2013tunneling,RevModPhys.87.1037,JorgeEPs, avila2019non} and odd-frequency pairing \cite{PhysRevB.105.094502}.  However, the realization of bulk Fermi arcs due to NH topology using only conventional superconductors still remains an open problem.

 \begin{figure}[!t]
\centering
	\includegraphics[width=0.9\columnwidth]{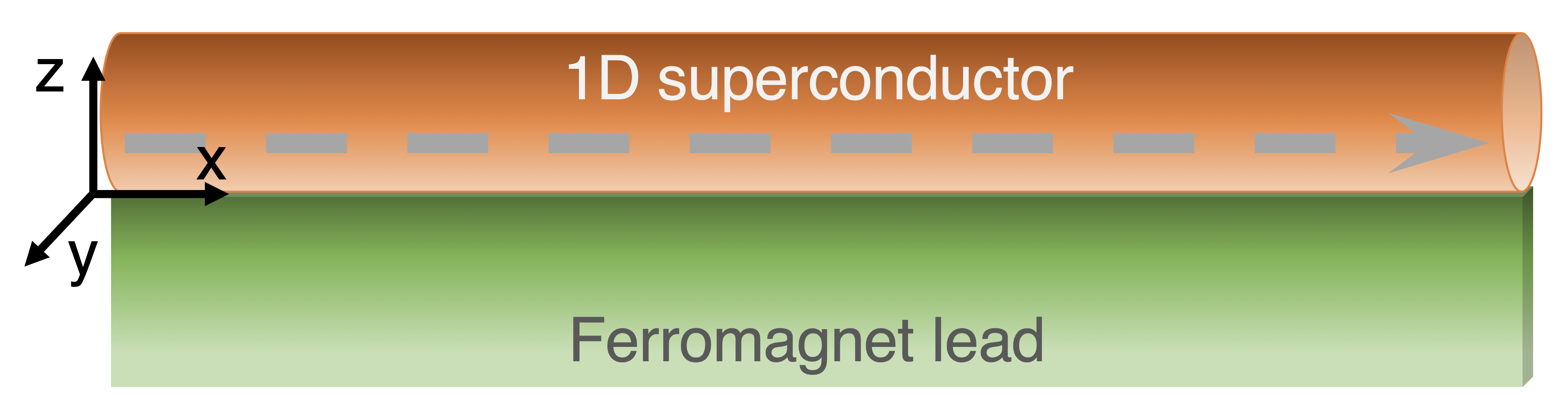}
	\caption{Sketch of a NH superconducting system: a 1D superconductor (orange)  subjected to a Zeeman field (dashed gray arrow) coupled to a ferromagnet lead (green).}
\label{Fig1} 
\end{figure}

In this work, we consider an experimentally feasible NH superconducting system and demonstrate the formation of stable bulk Fermi arcs, which are superconducting bulk excitations and thus we coin them \emph{bulk Bogoliubov Fermi arcs} (BBFAs). In particular, we engineer a realistic NH superconducting system by coupling a semi-infinite ferromagnet lead to a conventional superconductor with spin-orbit coupling, which is experimentally a feasible structure for superconductor-semiconductor \cite{Lutchyn2018Majorana, prada2019andreev, flensberg2021engineered} and ferromagnet systems \cite{liu2019semiconductor,vaitiekenas2021zero,escribano2022semiconductor,PhysRevB.105.L041304,razmadze2022supercurrent}. We discover the emergence of different types of EPs under an applied magnetic (Zeeman) field, marking the ends of regions with zero real energy and thus defining the formation of stable BBFAs. We further show that the BBFAs are naturally connected to the Hermitian topological phase transition, but importantly appear at lower Zeeman fields and in an overall larger parameter space. We also find that these unique bulk states dramatically increase the spectral weight, thus offering simple detection, e.g., using angle-resolved photoemission spectroscopy (ARPES). Our results reveal that the interplay of non-Hermiticity and conventional superconductivity gives rise to unexpected bulk phenomena that can considerably enhance the properties of superconducting systems.

%%%%%%%%%%%%%%%%%%%%%%%%%%%%%%%
% SECTION 2:       System: NH Rashba superconductor       %
%%%%%%%%%%%%%%%%%%%%%%%%%%%%%%%
\section{NH superconducting system}
\label{section2}
We consider an open system by coupling a conventional superconductor to a semi-infinite ferromagnet lead, as illustrated in Fig.\,\ref{Fig1}. This material junction is modelled by the effective NH Hamiltonian given by
\begin{equation}
\label{RashbaBdG1}
H_{\rm eff}=H_{\rm closed}^{\rm 1D}+\Sigma^{r}(\omega=0)\,, 
\end{equation}
where $H_{\rm closed}^{\rm 1D}$ describes the closed, i. e.~Hermitian, superconductor and $\Sigma^{r}(\omega=0)$ is a  spin-dependent retarded self-energy evaluated at zero frequency, appearing due to the coupling to the semi-infinite ferromagnet lead.  For $H_{\rm closed}^{\rm 1D}$ we consider 1D conventional spin-singlet $s$-wave superconductors with spin-orbit coupling (SOC), motivated both by them realizing (Hermitian) topological superconductivity \cite{PhysRevLett.105.077001,PhysRevLett.105.177002} and  because they describe superconductor-semiconductor systems within experimental reach  \cite{Lutchyn2018Majorana,prada2019andreev,flensberg2021engineered}. Similarly, the use of ferromagnets in these hybrid structures has also recently been demonstrated.
\cite{liu2019semiconductor,vaitiekenas2021zero,escribano2022semiconductor,PhysRevB.105.L041304,razmadze2022supercurrent}. Thus, we set $H_{\rm closed}^{\rm 1D}=\xi_{k}\tau_{z}+i\alpha k \sigma_{y}\tau_{z}+B\sigma_{x}\tau_{z}+\Delta\sigma_{y}\tau_{y}$, where $\xi_{k}=(\hbar^{2}k^{2}/2m-\mu)$ is the kinetic energy with $k$ along $x$, $\mu$ is the chemical potential, $\alpha$ is the Rashba SOC strength, and $\Delta$ characterizes the  spin-singlet $s$-wave superconducting order parameter. Without loss of generality, we assume $\alpha=1$, $m=1$, and $\hbar=1$. Also,  $B$ is a Zeeman field  from an applied magnetic field along the $x$-axis, inducing a (Hermitian) topological phase transition at $B_c=\pm \sqrt{\Delta^{2}+\mu^{2}}$ \cite{PhysRevLett.105.077001,PhysRevLett.105.177002}. 
 The self-energy $\Sigma^{r}$ can be obtained analytically 
\cite{PhysRevResearch.1.012003, PhysRevB.105.094502}, and is given in the wide band limit by  $\Sigma^{r}_{\rm e,h}(\omega=0)=-i \Gamma \sigma_{0}-i\gamma \sigma_{z}$ in both the electron ($e$) and hole ($h$) parts. Here,  $\Gamma=(\Gamma_{\uparrow}+\Gamma_{\downarrow})/2$ and $\gamma=(\Gamma_{\uparrow}-\Gamma_{\downarrow})/2$, where $\Gamma_{\sigma}=\pi|t'|^{2}\rho_{\rm L}^{\sigma}$ with $t'$ the hopping amplitude into the lead from the superconductor and $\rho_{\rm L}^{\sigma}$  being the surface  density of states of the lead for spin $\sigma=\uparrow$ and  $\downarrow$.  We finally note that this effective NH Hamiltonian  $H_{\rm eff}$ exhibits particle-hole symmetry given by $H(k)=-\hat{C}^{-1}H(-k)\hat{C}$, where $\hat{C}=\sigma_{0}\tau_{x}C$, with $C$ being the complex conjugation \cite{pikulin2012topological,PhysRevB.87.235421,ioselevich2013tunneling,RevModPhys.87.1037,JorgeEPs,avila2019non}. As we show below, this symmetry is important for understanding the properties of the system.

The pure imaginary (Im) self-energy renders $H_{\rm eff}$   in Eq.\,(\ref{RashbaBdG1}) NH \footnote{It is worth pointing out that non-Hermiticity can be also obtained from impurities, although earlier studies have shown that impurities produce self-energies that are different and more complicated than when using ferromagnet leads \cite{PhysRevB.99.165145,PhysRevB.104.035413}.}.
We are here particularly interested in the emergence of EPs and the formation of bulk Fermi arcs. For this purpose, we first obtain the eigenvalues
\begin{equation}
\label{NHRashbaEval}
E_{e_{\nu}(h_{\nu})}=-i\Gamma\pm\sqrt{A_{1}(k)+(-1)^{\nu}2\sqrt{A_{2}(k)}}\,,
\end{equation}
where $\nu=1(2)$ labels the first (second) $e$-like and $h$-like eigenvalues and
\begin{equation}
\label{A1A2}
\begin{split}
A_{1}&=\xi_{k}^{2}+\alpha^{2}k^{2}+B^{2}+\Delta^{2}-\gamma^{2}\,,\\
A_{2}&=\xi_{k}^{2}(B^{2}+\alpha^{2}k^{2})+B^{2}\Delta^{2}-\gamma^{2}\xi_{k}^{2}\,.
\end{split}
\end{equation}
The immediate observation from Eqs.\,(\ref{NHRashbaEval}-\ref{A1A2}) is that in the Hermitian regime, $\Gamma_{\uparrow,\downarrow}=0$,  the four eigenvalues in Eq.\,(\ref{NHRashbaEval}) are real (Re), but at any $\Gamma_{\uparrow,\downarrow}\neq0$, all four eigenvalues generally acquire an Im part reflecting the emergence of NH physics. Due to the retarded nature of the self-energy, these imaginary parts of the eigenvalues are expected to be negative. The inverse of this Im part characterizes the quasiparticle lifetime in the superconductor \cite{datta1997electronic}, thus giving a clear physical interpretation of  non-Hermiticity. For $\Gamma_{\uparrow}=\Gamma_{\downarrow}$, we get $\gamma=0$ and all the four eigenvalues share the same Im term, $-i\Gamma$.   The situation becomes  decisively more interesting when  $\Gamma_{\uparrow}\neq\Gamma_{\downarrow}$ as that gives rise to eigenvalues with different Im terms.  

Moreover, it is important to note that the particle-hole symmetry of $H_{\rm eff}$ implies  that, if $E_{i}$ is an eigenvalue, then $-E_{i}^{*}$ is also an eigenvalue.  Taking into account the four eigenvalues above, this symmetry requirement can be satisfied in two distinct ways. First, we can have pairs of eigenvalues $(E_{+},E_{-})$ with $E_{\pm}=\pm E -i \Gamma $ located symmetrically at opposite sides of the Im axis, where $E_{\pm}=-E^{*}_{\mp}$.  Here,  the $\pm$ subscript denote the electron and hole levels in Eq.\,(\ref{NHRashbaEval}), with $E$ being the (real-valued) energy and $\Gamma$ being the decay rate or lifetime of the quasiparticle. Second, it is also possible to have independent non-degenerate self-conjugate eigenvalues locate on the Im axis, where always $E_{\pm}=-E^{*}_{\pm}$. In this context, the bifurcation of two modes with $E_{\pm}=-E^{*}_{\mp}$ into two  modes with zero real energy and different lifetimes with $E_{\pm}=-E^{*}_{\pm}$ is a non-trivial effect that defines an EP in open systems. 
As a consequence, a non- degenerate pole $E=-i\Gamma$ on the imaginary axis   cannot acquire a nonzero real part without breaking the self-conjugation symmetry imposed by particle-hole symmetry \cite{pikulin2012topological,PhysRevB.87.235421,ioselevich2013tunneling,RevModPhys.87.1037,JorgeEPs,avila2019non}.
As we will see next, this, together with the impact of the couplings, has important consequences for the emergence of EPs and bulk Bogoliubov Fermi arcs in Eq.\,(\ref{RashbaBdG1}).

%%%%%%%%%%%%%%%%%%%%%%%%%%%%%%%
% SECTION 2:  Results: Second order Exceptional points     %
%%%%%%%%%%%%%%%%%%%%%%%%%%%%%%%
\section{EPs and bulk Bogoliubov Fermi arcs}
\label{section3}
To  begin, we study the emergence of second order EPs, appearing when two eigenvalues, and their respective eigenvectors, coalesce \cite{heiss2004exceptional, Heiss_2012}.  
This means searching for regimes where two eigenvalues merge and their eigenvectors become parallel. By analyzing Eq.\,(\ref{NHRashbaEval}) and the behavior of $A_{1,2}$, we realize that there are two possibilities for second order EPs. First, the two $E_{e_{\nu}} (E_{h_{\nu}})$ eigenvalues can coalesce when $A_{1}>0$ and $ 2\sqrt{A_{2}} =0$, giving rise to EPs with finite Re energies. These EPs need finite SOC and $\gamma$, but notably do not require superconductivity, but are tied to properties of the normal state. We thus do not focus on them here
\footnote{These EPs also do not require a Zeeman field $B$ and have recently been discussed in Ref.\,\cite{Cayao2023}. The momenta at which they appear are found by solving $2\sqrt{A_{2}}=0$, which at $B=0$ leads to $k_{\rm EP}=\pm|\gamma|/\alpha$. These EPs remain robust at weak $B$ but then merge and split as $B$ increases. We also find no bulk Fermi arcs generated by these EPs.}. 
The other possibility for EPs is more interesting: the $\nu=1$ electron ($E_{e_{1}}$) and  hole ($E_{h_{1}}$) eigenvalues coalesce for $k$-values where  ${\rm Re}(2\sqrt{A_{2}})=A_{1}>0$ and ${\rm Im}(2\sqrt{A_{2}})=0$, which leads to ${\rm Re}E_{j_{1}}=0$ and ${\rm Im}E_{j_{1}}\neq0$, $j=e,h$. Remarkably, the latter two conditions also guarantee an existence of bulk Fermi arcs \cite{kozii2017non, science359Zhou, doi:10.7566/JPSCP.30.011098, PhysRevLett.125.227204, PhysRevLett.127.186601}. 
However, unlike the bulk Fermi arcs earlier only found in normal-state systems \cite{kozii2017non,science359Zhou,doi:10.7566/JPSCP.30.011098, PhysRevLett.125.227204, PhysRevLett.127.186601}, here they consist of superconducting, or Bogoliubov, quasiparticle excitations, and we therefore name them \textit{bulk Bogoliubov Fermi arcs} (BBFAs).

The stability of the EPs and BBFAs is ensured by the underlying particle-hole symmetry of  $H_{\rm eff}$ discussed in Section \ref{section2}. Specifically, particle-hole symmetry enables a robustness of EPs formed by two  levels with zero Re energies and different lifetimes obeying self-conjugation symmetry $E_{\pm}=-E^{*}_{\pm}$, which is the case for the second order EPs discussed above. In this situation it is natural to define a NH spectral topological invariant $N$ that counts the number of eigenvalues with zero Re energies. This definition links to the invariant used in particle-hole symmetric NH systems \cite{PhysRevX.8.031079,PhysRevB.99.041202} and also connects to the classification of topology of S-matrices \cite{pikulin2012topological,PhysRevB.87.235421,ioselevich2013tunneling,RevModPhys.87.1037,JorgeEPs,avila2019non}. In detail, for two particle-hole symmetric eigenvalues, before the EP, both have finite Re and Im parts, 
leading to a topologically trivial gapped phase with $N=0$. At the EP, the two eigenvalues merge at zero Re energy, generating $N=2$ and signaling a transition from a gapped to a gapless phase that is topologically non-trivial. Within the BBFA region, the number $N=2$ does not change, revealing that BBFAs represent a robust and  NH bulk topological phenomenon 
\footnote{The topological properties of EPs can also be discussed in terms of the complex gaps of the spectrum, see e.g., \cite{PhysRevLett.123.066405}, where topological stability of the EP is ensured by a point gap, revealed when encircling the EP. When BBFAs emerge, a real line gap is closed, while an imaginary gap remains finite over the extent of the BBFAs. It is this imaginary line gap that protects the BBFAs in the language of the complex energy gap topology \cite{PhysRevLett.123.066405}.}.

\begin{figure}[!t]
\centering
	\includegraphics[width=0.99\columnwidth]{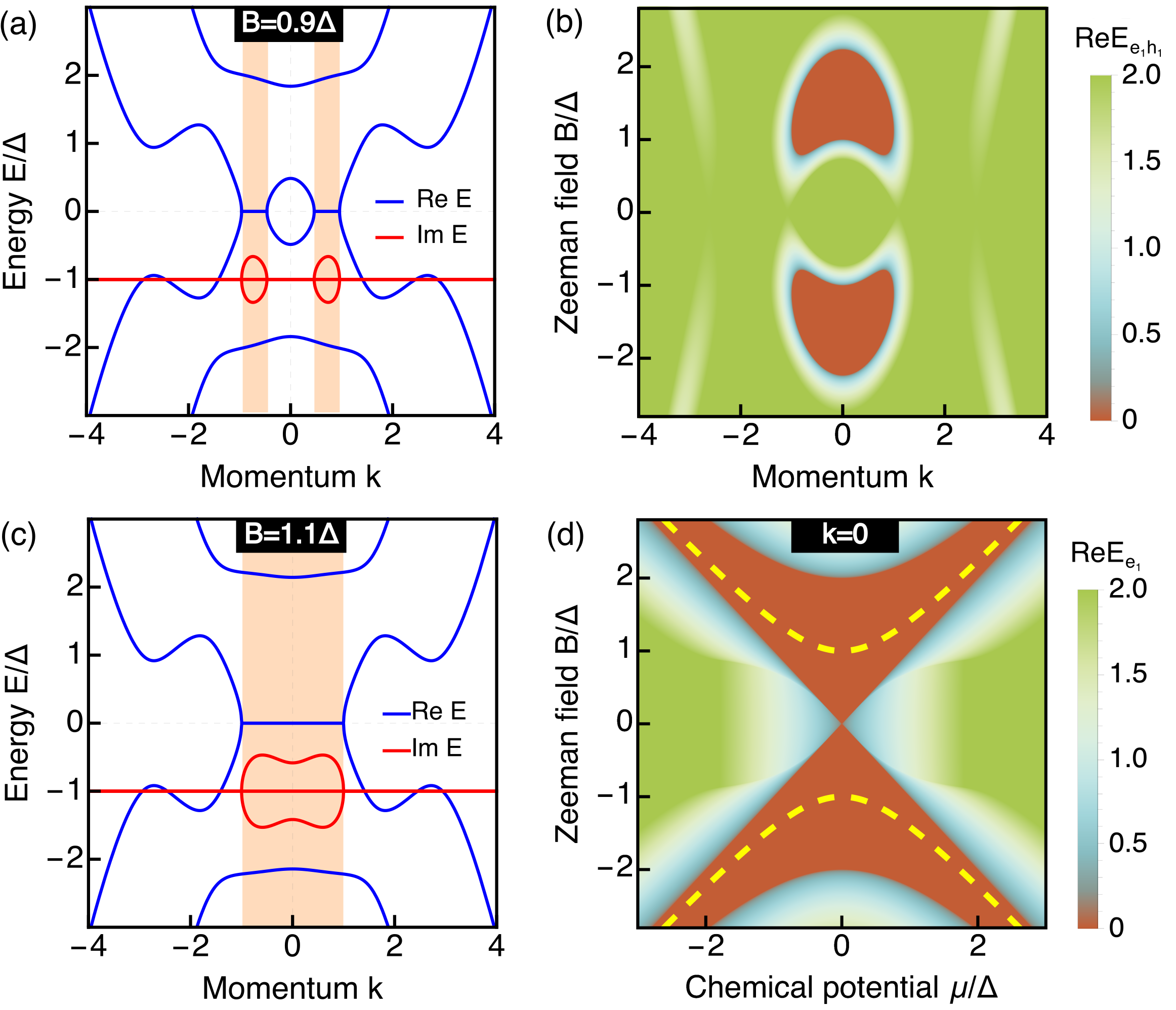}
	\caption{EPs and bulk Bogoliubov Fermi arcs: (a,c) Re (blue) and Im (red) parts of eigenvalues as a function of momentum $k$ at two different magnetic fields $B$.  Shaded orange regions indicate zero real energies (BBFAs) with their borders marking the EPs. (b) ${\rm Re}E_{e_{1}h_{1}}={\rm Re}(E_{e_{1}}-E_{h_{1}})$ as a function of $k$ and $B$, with orange regions indicating ${\rm Re}E_{e_{1}(h_{1})}=0$.  (d) ${\rm Re}E_{e_{1}}$ at $k=0$ as a function of $B$ and the chemical potential $\mu$, with the orange region indicating ${\rm Re}E_{e_{1}}=0$. Yellow dashed lines depict the Hermitian TPT at $B_c=\pm\sqrt{\Delta^{2}+\mu^{2}}$. Parameters: $\mu=\Delta$, $\Gamma_{\uparrow}=2\Delta$, $\Gamma_{\downarrow}=0$.   
	}
\label{Fig3} 
\end{figure}

The momenta $k_{\rm EP}$ for the above EPs can be obtained by solving $A_{1}={\rm Re}(2\sqrt{A_{2}})$, as those also satisfy ${\rm Im}(2\sqrt{A_{2}})=0$. We also verify that the corresponding eigenvectors become parallel at the EPs, see Appendix \ref{Appendix1}. These EPs generally requires the interplay of  non-Hermiticity (here $\gamma\neq0$), Zeeman field, and superconductivity, reflecting a distinctly superconducting origin. 
For vanishingly small SOC, the EPs appear at very low Zeeman fields ($B\ll\Delta$) when $\gamma\sim\Delta$. In this case, we find $\pm k^{\pm}_{\rm EP}=\pm\sqrt{k_{\rm F}^{2}\pm \kappa^{2}}$, where $k_{\rm F}=\sqrt{2m\mu/\hbar^{2}}$ and $\kappa=\sqrt{(2m/\hbar^{2})\sqrt{B^{2}-(\Delta-\gamma)^{2}}}$. Assuming $\mu\neq0$, $k_{\rm F}>\kappa$ implies four different real EP momenta, but is otherwise reduced to two.  
For stronger SOC, $B\sim\Delta$ is needed for similar values of $\gamma$ to find EPs. To visualize these EPs, we plot in Fig.\,\ref{Fig3}(a,c) the eigenvalues as a function of  $k$ at two finite $B$, with the Re and Im parts shown in blue and red, respectively. We observe that the Re parts of the lowest two levels vanish for a range of negative and positive $k$, while the Im parts split up, indicated by shaded orange regions. 
These orange regions, with ${\rm Re}E=0$ and ${\rm Im}E\neq0$, define the topologically protected BBFAs.
The end points of the BBFAs are at the EP momenta, where both eigenvalues and eigenvectors fully coalesce. Analytically, we find the EPs when $A_{1}={\rm Re}(2\sqrt{A_{2}})$, while the BBFAs satisfy the less stringent condition $A_{1} \leq {\rm Re}(2\sqrt{A_{2}})$. By increasing $B$, the inner EPs (closer to $k=0$) merge, giving rise to a single, longer, BBFA connecting the outer EPs, see Fig.\,\ref{Fig3}(c). 
 To support the generic occurrence of EPs and BBFAs, we plot in Fig.\,\ref{Fig3}(b) the property ${\rm Re}E_{e_{1}h_{1}}={\rm Re}(E_{e_{1}}-E_{h_{1}})$ as a function of $k$ and $B$. Here, the orange region indicates ${\rm Re}E_{e_{1}h_{1}}=0$, thus revealing that the formation of BBFAs occurs in a large parameter region that is also highly tunable by the external field $B$. 
 We have also verified that  for small $\mu$,  which is commonly controlled by doping or gating \cite{datta1997electronic}, the EPs and BBFAs require very low Zeeman fields ($B\ll\Delta$), even for strong SOC.  We also stress that, while SOC is not needed for these NH phenomena, its finite value shapes the profile of the  spectrum such that weaker SOC enables larger separation of outer EPs, thus favoring the formation of longer BBFAs.

The BBFAs in Fig.\,\ref{Fig3} have an interesting connection to the topological phase transition (TPT) in the Hermitian regime of the system, occurring at the Zeeman field $B_c=\pm \sqrt{\mu^{2}+\Delta^{2}}$. In fact, the vanishing of  ${\rm Re}E$ at the BBFA is highly reminiscent of the Hermitian TPT, as both close the energy gap. However, the TPT occurs at a single $k$ and only for $B\geq \Delta$.  To  illustrate the situation, we plot in Fig.\,\ref{Fig3}(d)  the property ${\rm Re}E_{e_{1}}$ as a function of $\mu$ and $B$, with the orange region indicating ${\rm Re}E_{e_{1}}=0$. We also mark the Hermitian TPT with yellow dashed lines.  The main feature we note is that, while the gap closing in the Hermitian regime (TPT) occurs along a single line in parameter space, the gap closing in the NH case (BBFA) occurs in a much larger parameter regime and also at much lower $B$. To provide further insight, we can study the lowest positive level at $\mu=0$ and $k=0$: $E_{e_{1}}=-i\Gamma+\sqrt{(B-\Delta)^{2}-\gamma^{2}}$. Here, the gap closes, ${\rm Re}E_{e_{1}} = 0$, at $B=\Delta\pm\gamma$, which makes it evident that it is the non-Hermiticity causing the substantial lower $B$ compared to the Hermitian TPT at $B_c = \Delta$. We can thus view the topologically protected BBFA as the natural NH extension of the Hermitian TPT, while at the same time being much more prevalent in parameter space than the TPT. This may lead to advantageous effects, e.g.~ allowing realization of Majorana zero modes at  Zeeman fields lower than those in purely Hermitian systems.

%%%%%%%%%%%%%%%%%%%%%%%%%%%%%%%
% SECTION 3:  Results: Two-fold 2nd order Exceptional points     %
%%%%%%%%%%%%%%%%%%%%%%%%%%%%%%%
\section{Two-fold second order EPs}
\label{section4}
Having established the existence of EPs with BBFAs, we next identify another interesting type of EPs in superconducting systems: two-fold second order EPs, which occur when the Re part of all four eigenvalues coalesce, but their Im parts merge into two distinct values.  Although this might naively be viewed as a fourth order EP,  their distinct Im parts  restrict them from being of fourth order and it is more appropriate to refer to them as two-fold second order EPs. 
As before, the EPs are determined by the behavior of $A_{1,2}$ in Eq.\,(\ref{NHRashbaEval}). In fact, we find that  when  $A_{1}\leq0$,  $0 \leq |2{\rm Re}\sqrt{A_{2}}|\leq|A_{1}|$, and ${\rm Im}\sqrt{A_{2}}=0$, we find two-fold second order EPs for the strict equality and BBFAs for the inequality, since the Re part of all four eigenvalues ($E_{e_\nu}$ and $E_{h_\nu}$) coalesce at zero energy.
The EP momenta can be found by solving $A_{2}=0$, acquiring complicated expressions, but which reduce to $\pm k_{\rm EP_{*}}^{\pm}=\pm \sqrt{k_{\rm F}^{2}\pm \kappa_{*}^{2}}$  for vanishing SOC, where $\kappa_{*}=\sqrt{(2mB\Delta)/(\hbar^{2}\sqrt{\gamma^{2}-B^{2}})}$. For $k_{\rm F}>\kappa_{*}$, the system exhibits four real EP momenta while only two exists when $k_{\rm F}<\kappa_{*}$, leading to a strong control by $\mu$.  As expected, the eigenvectors associated with the eigenvalues also coalesce at the two-fold EPs, see Appendix \ref{Appendix2}. The EP conditions may naively seem hard to achieve, but we verify that, although they need stronger non-Hermiticity ($\gamma>\Delta$) than the second order EPs in the previous section, they actually appear at weaker Zeeman fields ($0<B\leq\Delta$).

\begin{figure}[!t]
\centering
	\includegraphics[width=0.99\columnwidth]{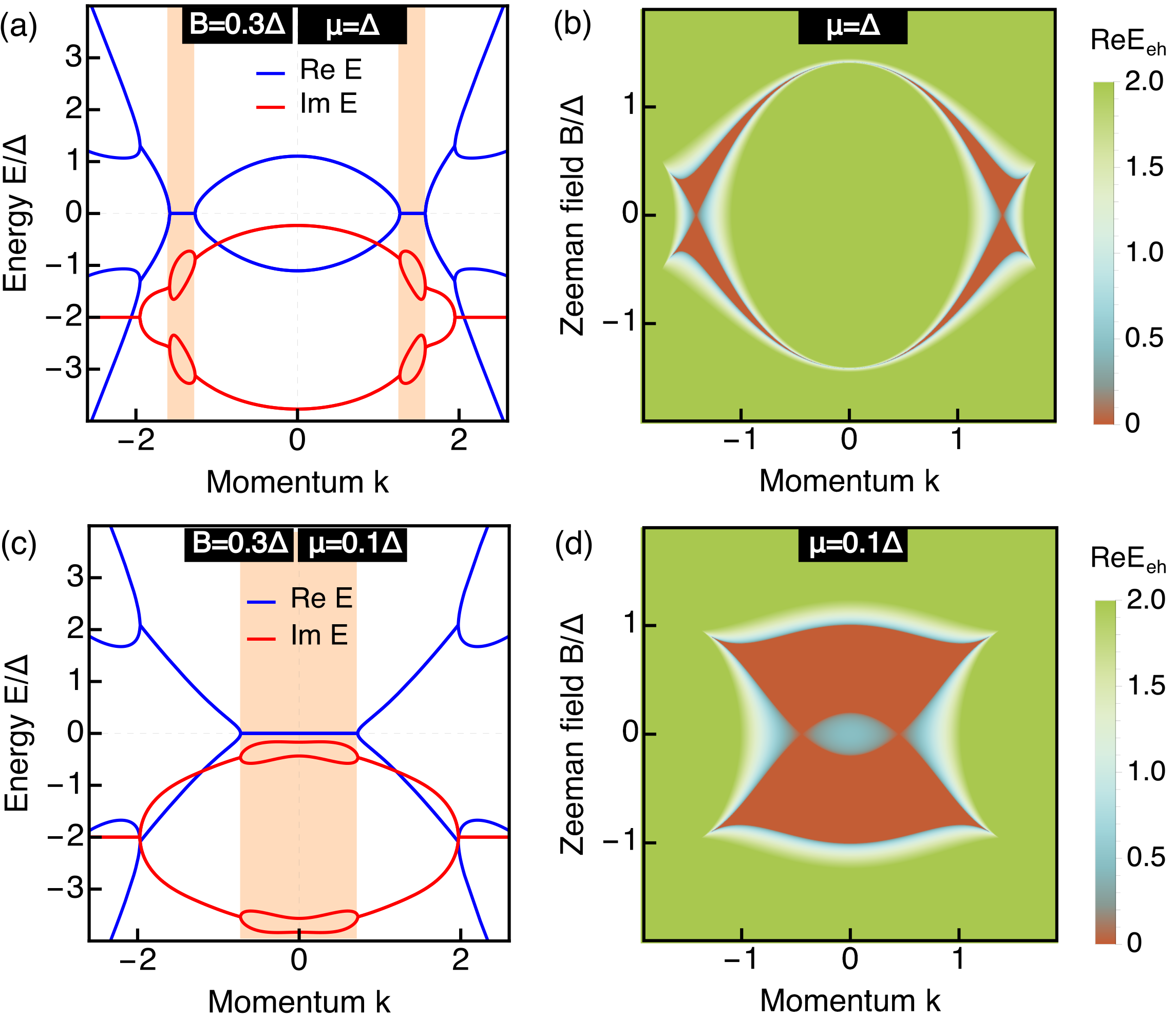}
	\caption{Two-fold second order EPs: 
	(a) Re (blue) and Im (red)  parts of eigenvalues as a function of momentum $k$ for $\mu=\Delta$ and $B=0.3\Delta$. Shaded orange regions indicate zero real energies (BBFAs) with their borders marking the EPs.   (b) ${\rm Re}(E_{eh})={\rm Re}(E_{e_{1}}+E_{e_{2}}-E_{h_{1}}-E_{h_{1}})$ as a function of $B$ and $k$ for $\mu=\Delta$ and $\Gamma_{\uparrow}=4\Delta$, with orange regions indicating ${\rm Re}(E_{eh}) = 0$. (c,d) Same as (a,b) but for $\mu=0.1\Delta$. Parameters:  $\Gamma_{\uparrow}=4\Delta$, $\Gamma_{\downarrow}=0$.   }
\label{Fig4} 
\end{figure}

 To visualize the two-fold second order EPs, we plot in Figs.\,\ref{Fig4}(a,c)  the eigenvalues at large $\gamma$, weak $B$, and two different values of chemical potential set by $\mu$ with the Re and Im parts depicted in blue and red, respectively. At high $\mu$, the Re part of all four eigenvalues merge at zero energy for two small ranges of positive and negative $k$ forming BBFAs, while their Im parts split and form lobes within these regimes at two different values, marked as orange regions in Fig.\,\ref{Fig4}(a). These orange BBFA regions are terminated by the EP momenta $\pm k_{\rm EP_{*}}^{\pm}$. The formation of BBFAs can be also seen in  Fig.\,\ref{Fig4}(b) where we plot  ${\rm Re}(E_{eh})={\rm Re}(E_{e_{1}}+E_{e_{2}}-E_{h_{1}}-E_{h_{1}})$, which clearly illustrates the appearance of BBFAs for all $B<\Delta$, see orange regions. At low $\mu$, Figs.\,4(c,d) show how the inner EPs ($\pm k^{-}_{\rm EP_{*}}$) merge and leave a much longer zero-energy line now connecting the outer EPs ($\pm k^{+}_{\rm EP_{*}}$), thus giving rise to a much more extended BBFAs. 

An interesting property of the BBFAs from the two-fold second order EPs  in Fig.\,\ref{Fig4}(a,c), is that they are distinct  from the BBFA from the second order EPs in Fig.\,\ref{Fig3}(a,c).
This can be seen by noting that the Im parts associated with the two types of BBFAs exhibit a different structure. This then implies that the lifetimes, defined by the inverse of the Im part of eigenvalues, acquire distinctly different behavior for the  two BBFAs. 
Moreover, we note that the two-fold second order EPs  need $B$ values much lower than those of the second order EPs, appearing even for $B\ll \Delta$, which makes them highly tunable with a small external magnetic field, even much smaller than required to reach the Hermitian TPT.

%%%%%%%%%%%%%%%%%%%%%%%%%%%%%%%
%                       SPECTRAL FUNCTION                               %
%%%%%%%%%%%%%%%%%%%%%%%%%%%%%%%
\section{Spectral signatures}
Having demonstrated the emergence of two different types of EPs and BBFAs, we finally show that these NH effects can be directly probed  by the spectral function 
$A(k,\omega)=-{\rm Im Tr} {(G^{r}-G^{a})}$, where $G^{r}=(\omega-H_{\rm eff})^{-1}$  and $G^{a}=(G^{r})^{\dagger}$ represent the retarded and advanced Green's functions, respectively \cite{mahan2013many,zagoskin}. This approach to obtain $A(k,\omega)$ is common and directly reveals the power of Green's functions. Furthermore, we point out that the spectral function is a measurable observable that can be accessed by tools such as ARPES \cite{hufner2013photoelectron,lv2019angle,yu2020relevance,doi:10.7566/JPSJ.84.072001,RevModPhys.93.025006,PhysRevResearch.4.L022018}.

\begin{figure}[!t]
\centering
	\includegraphics[width=0.99\columnwidth]{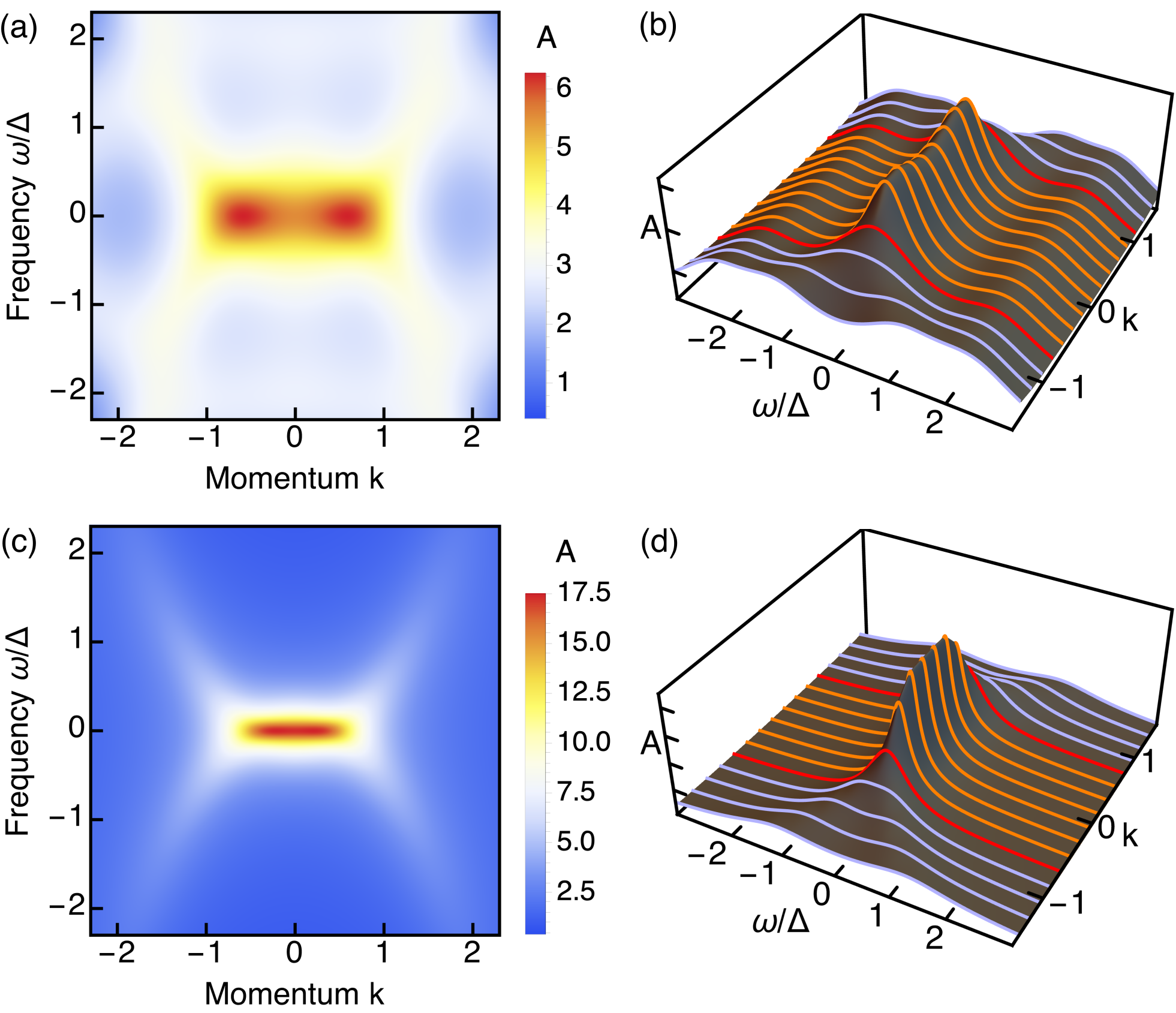}
	\caption{Spectral signatures: Spectral function $A(\omega,k)$ as a function of frequency $\omega$ and  $k$ with clear features of the BBFAs from  second order (a,b)  and two-fold second order (c,d) EPs. Line cuts of $A$  (b,d) with EPs (red), BBFAs (orange) and the region beyond (light blue). In (a,b) $B=1.1\Delta$, $\mu=\Delta$, and $\Gamma_{\uparrow}=2\Delta$, while in (c,d) $B=0.3\Delta$, $\mu=0.1\Delta$, and $\Gamma_{\uparrow}=4\Delta$.  Remaining parameters:  $\Gamma_{\downarrow}=0$.}
\label{Fig5} 
\end{figure}

To visualize the behavior of the spectral function, we plot $A$ in Fig.\,\ref{Fig5}(a,c) as a function of frequency $\omega$ and $k$ for parameters generally  favoring the formation of BBFAs and EPs found  in the previous two sections, respectively.  In Fig.\,\ref{Fig5}(b,d) we extract line cuts for distinct $k$-values spanning the EP momenta.
We directly observe the emergence of high intensity regions at $\omega=0$ for a line of momentum values bounded by the EP momenta, thus directly revealing the formation of the BBFAs. Even though both BBFAs produce quite similar spectral weight signatures, they exhibit subtle and interesting differences tied to the different structures of their eigenvalues. This can be understood by noting that $A$ is a sum of Lorentzians  centered at ${\rm Re}E_{j_{\nu}}$, with their height and width characterized by ${\rm Im}E_{j_{\nu}}$ \cite{kozii2017non}. As a consequence, at the EPs, marked by red lines in Fig.\,\ref{Fig5}(b,d), the spectral function undergoes a transition along $k$ from having two or more resonances to having a single resonance centered at $\omega=0$.
The $\omega=0$ resonances between the red curves, depicted by orange lines in Fig.\,\ref{Fig5}(b,d) and directly revealing the BBFA, are notably higher and narrower for the two-fold second-order EP system in (d) compared to the second-order EP system in (b). This is due to the distinctly different Im parts of the eigenvalues, or lifetimes, of the two different types of EPs. More coalescence among the Im parts, as for the two-fold second order EPs, results in higher and narrower spectral peaks also for its accompanying BBFAs. This explicitly illustrates how the properties of the EPs determine the physical manifestation of the BBFAs.

%%%%%%%%%%%%%%%%%%%%%%%%%%%%%%%
%                           CONCLUSIONS                                        %
%%%%%%%%%%%%%%%%%%%%%%%%%%%%%%%

\section{Conclusions}
\label{sec:Concl}
In summary, we show the formation of topologically protected bulk Fermi arcs in  realistic NH superconducting systems and coin them BBFAs due to their superconducting nature. These BBFAs are zero-energy bulk states, unlike the Majorana zero-energy modes appearing as boundary states in Hermitian topological superconductors. Still, we are able to establish as a byproduct a connection between  BBFAs and the Hermitian topological phase transition,  obtaining that the BBFAs are much more prevalent in parameter space and also are highly tunable by  both the chemical potential and  the external magnetic field.  Interestingly, unlike Hermitian topological superconductors, the NH superconductor requires lower Zeeman fields to undergo the topological phase transition. We also reveal that the BBFAs induce large spectral features which can be directly detected by ARPES.  

Furthermore, it is important to point out that  systems similar to those   modelled here have already been fabricated using e.g.~Al/InAs superconductor-semiconductor hybrid structures \cite{krogstrup2015epitaxy, chang2015hard, PhysRevB.93.155402, Lutchyn2018Majorana, prada2019andreev, flensberg2021engineered}, showing both good proximity-induced superconductivity and controllable couplings to leads, placing our findings clearly within experimental reach.  Along these lines, we also stress that ferromagnets in such hybrid systems are currently being studied \cite{liu2019semiconductor,vaitiekenas2021zero,escribano2022semiconductor,PhysRevB.105.L041304,razmadze2022supercurrent}, and, despite challenges, this provides promising results supporting the feasibility of our proposal.  Our results can also be extended to higher dimensions, resulting in e.g.~bulk Bogoliubov Fermi surfaces and volumes. By uncovering unique phenomena emerging from the interplay of non-Hermiticity and conventional superconductivity, our work opens the route for greatly enhancing functionalities of superconducting hybrid structures.

%%%%%%%%%%%%%%%%%%%%%%%%%%%%%%%
%                        ACKNOWLEDGMENTS                               %
%%%%%%%%%%%%%%%%%%%%%%%%%%%%%%%
\begin{acknowledgements}
We thank  R.~Aguado, E.~J.~Bergholtz, and T. Yoshida for insightful discussions.  We acknowledge financial support from the Swedish Research Council  (Vetenskapsr\aa det Grants No.~2018-03488 and No.~2021-04121), the G\"{o}ran Gustafsson Foundation (Grant No.~2216),   the Knut and Alice Wallenberg Foundation, the  Letterstedt travel scholarship Foundation, and 
the Royal Swedish Academy of Sciences (Grant No.~PH2018-0006).
\end{acknowledgements}
%%%%%%%%%%%%%%%%%%%%%%%%%%%%%%%
%                                  APPENDIX                                  %
%%%%%%%%%%%%%%%%%%%%%%%%%%%%%%%

%%Only %% useful for appendix
%\cleardoublepage
%%\onecolumngrid
\appendix
\renewcommand{\thepage}{A\arabic{page}}
\setcounter{page}{1}
\renewcommand{\thefigure}{A\arabic{figure}}
\setcounter{figure}{0}

\section{Coalescence of eigenfunctions at EPs}
\label{Appendix}
In this work we study EPs in NH superconductors with SOC, described by the effective Hamiltonian $H_{\rm eff}$ in Eq.\,(\ref{RashbaBdG1}). For this reason in the main text we present the complex spectrum showing the coalescence of eigenvalues at EPs. To further support the nature of these EPs, here we show
 that the eigenfunctions at the identified EPs become parallel, namely, they  coalesce, as expected.  It is important to note that, even though the matrix structure of the effective Hamiltonian $H_{\rm eff}$ given by Eq.\,(\ref{RashbaBdG1}) is not complicated, the resulting expressions for the eigenfunctions are lengthy and not particularly enlightening, which motivates a numerical solution. The Hamiltonian $H_{\rm eff}$ has four eigenvalues $\{E_{e_1},E_{e_2},E_{h_1},E_{h_2}\}$ given by Eqs.\,(\ref{NHRashbaEval}) and four eigenvectors  $\{\Psi_{e_1},\Psi_{e_2},\Psi_{h_1},\Psi_{h_2}\}$ which we below obtain for the two types of EPs discussed in Sections \ref{section3} and \ref{section4}, respectively.

\subsection{Coalescence of the two lowest eigenfunctions}
\label{Appendix1}
\begin{figure}[!t]
\centering
	\includegraphics[width=0.99\columnwidth]{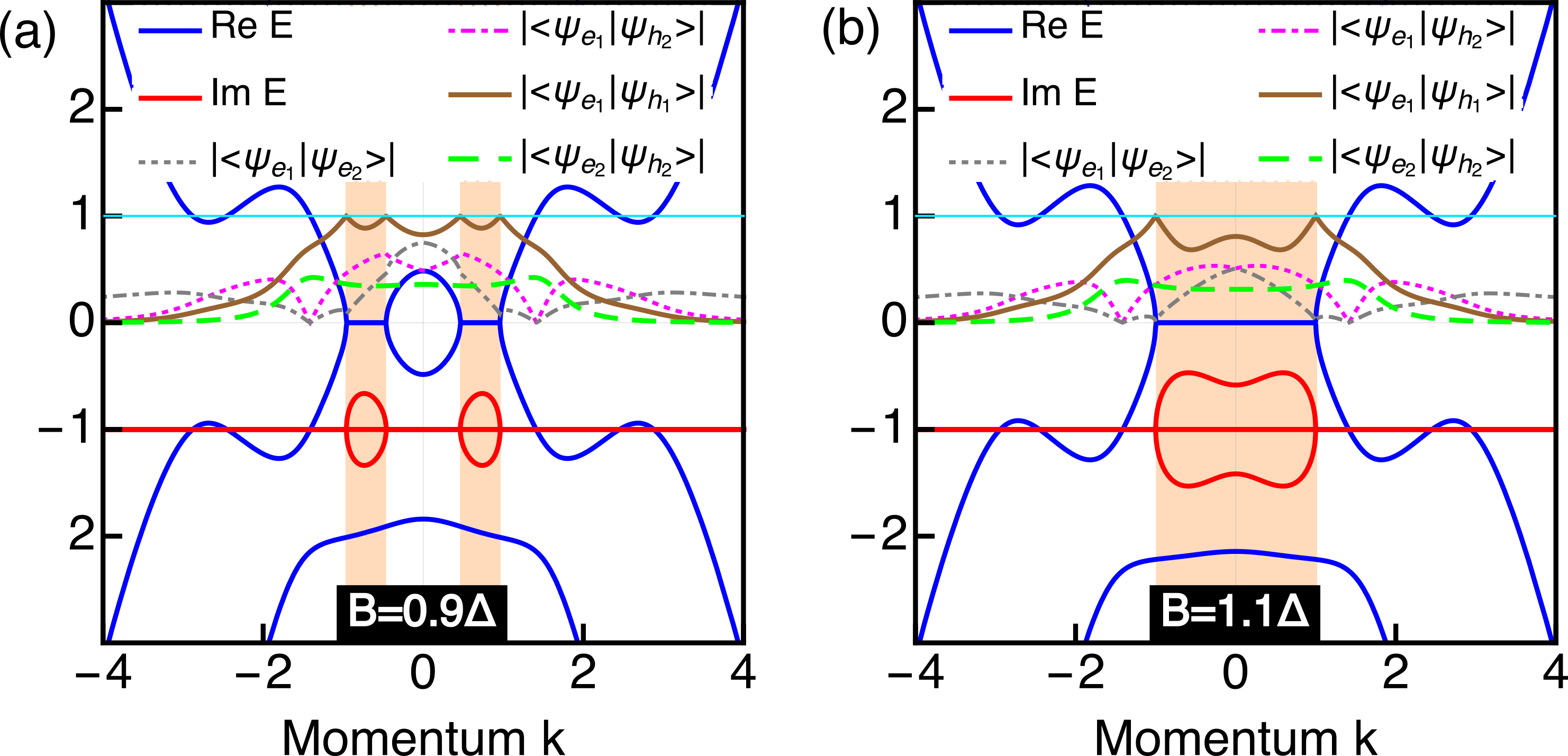}
	\caption{Second order EPs: (a) Re (blue) and Im (red)  parts of eigenvalues as a function of momentum $k$ for  $B=0.9\Delta$ (a) and $B=1.1\Delta$ (b). Shaded orange regions indicate zero real energies (BBFAs) with their borders marking the EPs.  Dotted gray, dashed-dot magenta, brown, and green curves show the absolute value of the eigenfunction overlaps  $|\bra{\psi_{e_1}}\ket{\psi_{e_{2}}}|$ and $|\bra{\psi_{e_\nu}}\ket{\psi_{h_\nu'}}|$, with $\nu,\nu'=1,2$, while cyan curves show $\bra{\psi_{e_{\nu}(h_{\nu})}}\ket{\psi_{e_\nu(h_{\nu})}}=1$. Parameters as in Fig.\,\ref{Fig3}.
}
\label{Fig1App} 
\end{figure}

To identify the EPs when the two lowest levels coalesce,  we need to find the momenta at which they occur following the conditions discussed  in  Section \ref{section3}. In particular, we need to solve  $A_{1}=2\sqrt{A_{2}}$ and find real values of $k$  that allow $A_{1}>0$ and ${\rm Im}(2\sqrt{A_2})=0$. We thus find four EP momenta, denoted here as $\pm k_{\rm EP}^{\pm}$, whose expressions are rather complicated but for zero SOC reduce to $\pm k^{\pm}_{\rm EP}=\pm\sqrt{k_{\rm F}^{2}\pm \kappa^{2}}$, where $k_{\rm F}=\sqrt{2m\mu/\hbar^{2}}$ and $\kappa=\sqrt{(2m/\hbar^{2})\sqrt{B^{2}-(\Delta-\gamma)^{2}}}$, as discussed in Section \ref{section3}. We find that at $\pm k_{\rm EP}^{\pm}$ the two lowest levels $\{E_{e_1},E_{h_1}\}$ and their respective eigenfunctions $\{\Psi_{e_1},\Psi_{h_1}\}$ coalesce, 
 \begin{align}
 {\rm Re}E_{e_1}&\equiv  {\rm Re}E_{h_1}=0\,,\\  
 {\rm Im}E_{e_1}&\equiv  {\rm Im}E_{h_1}\neq 0\,,\\
  \Psi_{e_1}&\equiv \Psi_{h_1}\,.
 \end{align}
We visualize this coalescence in Fig.\,\ref{Fig1App}, where we repeat 
Fig.\,\ref{Fig3}(a,c) now with additional curves for the eigenvector overlaps $\langle \Psi_{e_{1}}|\Psi_{e_{2}}\rangle$, $\langle \Psi_{e_{1}}|\Psi_{h_{2}}\rangle$, $\langle \Psi_{e_{1}}|\Psi_{h_{1}}\rangle$, and $\langle \Psi_{e_{2}}|\Psi_{h_{2}}\rangle$. At the EPs, marked by the borders of the orange regions, the overlap involving $\Psi_{e_{1}}$ and $\Psi_{h_{1}}$ reaches one and we have also verified that here $\Psi_{e_{1}} =\Psi_{h_{1}}$. Thus, the overlap reaching one reveals that these eigenvectors coalesce and thus become parallel.  Therefore, both the eigenvalues $\{E_{e_1},E_{h_1}\}$ and their eigenvectors $\{\Psi_{e_1},\Psi_{h_1}\}$ coalesce at the EPs, as discussed in Section \ref{section3}.

\subsection{Coalescence of all four eigenfunctions}
\label{Appendix2}
In the case of the two-fold EPs, we again follow the conditions for finding the EP momenta discussed in Section \ref{section4}. Thus, we solve $A_{2}=0$ for $k$ and then choose the ones that provide $A_{1}<0$. Under these conditions, we find four momenta $\pm k_{\rm EP_*}^{\pm}$, two   positive and two  negative, which at zero SOC reduce to the simpler form $\pm k_{\rm EP_{*}}^{\pm}=\pm \sqrt{k_{\rm F}^{2}\pm \kappa_{*}^{2}}$, where $\kappa_{*}=\sqrt{(2mB\Delta)/(\hbar^{2}\sqrt{\gamma^{2}-B^{2}})}$. 
It is worth noting that $\pm k_{\rm EP_*}^{\pm}$ generally appear around Fermi momenta $\pm k_{\rm F}$, even become equal to $\pm k_{\rm F}$ at zero Zeeman field. Beyond these simplifying limits, we have numerically verified that , at $\pm k_{\rm EP_*}^{\pm}$, the four eigenvalues and their respective eigenvectors coalesce in pairs:   the pair of eigenvalues $\{E_{e_\nu},E_{h_\nu}\}$  and its eigenvectors $\{\Psi_{e_\nu},\Psi_{h_\nu}\}$ with $\nu=1,2$ merge into a single value or vector.  More precisely, the following relations hold,
 \begin{equation}
 \begin{split}
 {\rm Re}E_{e_\nu}&\equiv  {\rm Re}E_{h_\nu}=0\,,\\  
 {\rm Im}E_{e_\nu}&\equiv  {\rm Im}E_{h_\nu}\neq0 \\
  \Psi_{e_\nu}&\equiv \Psi_{h_\nu}\,.
 \end{split}
 \end{equation}
 We note that, at these EPs $\pm k_{\rm EP_*}^{\pm}$, each pair merges into the same zero real energy but their imaginary parts and eigenvectors remain distinct between the two pairs. This is  the reason why we refer to these EPs as two-fold second order EPs and not a fourth order EP.     

\begin{figure}[!ht]
\centering
	\includegraphics[width=0.99\columnwidth]{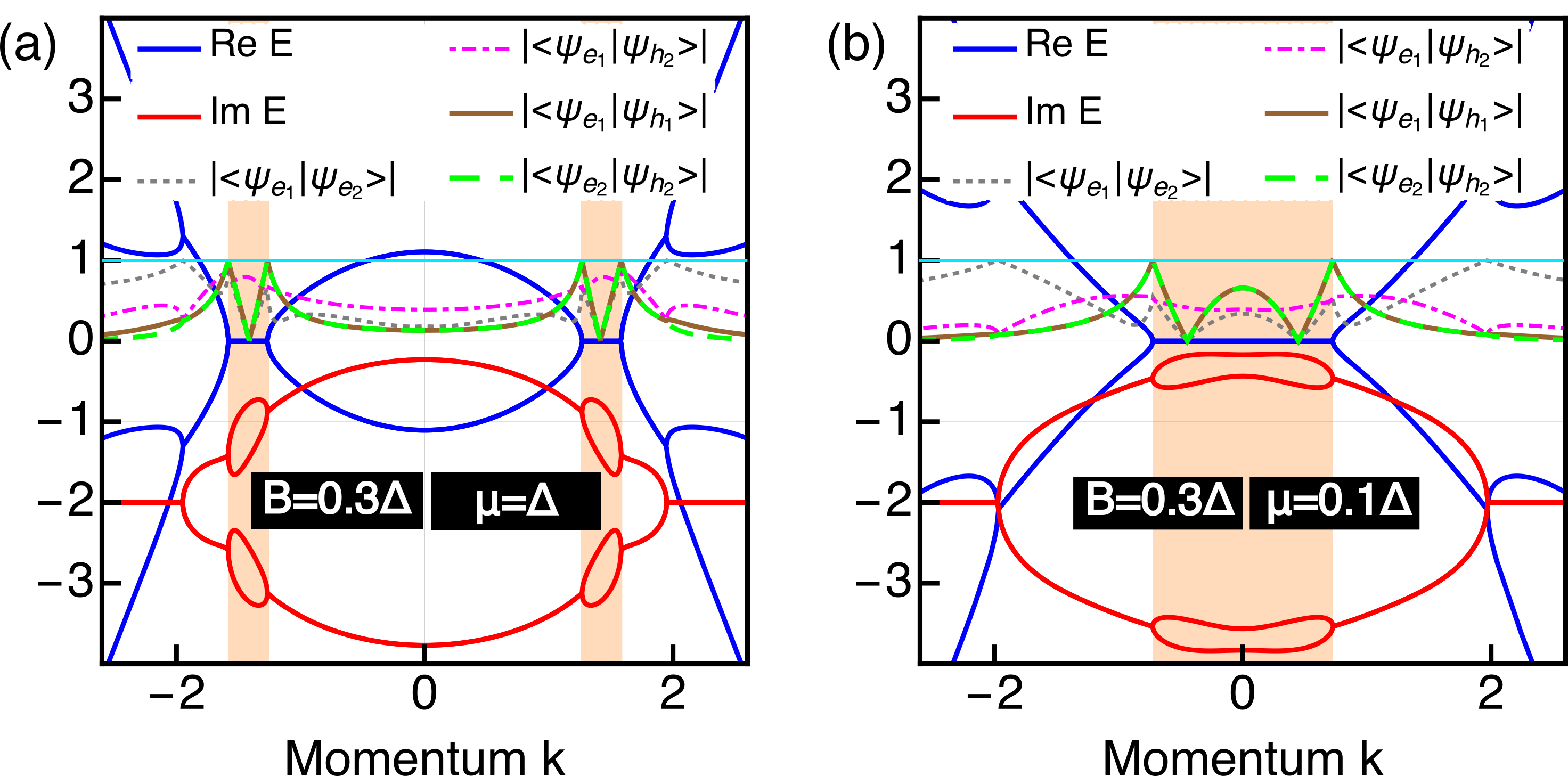}
	\caption{Two-fold second order EPs: (a) Re (blue) and Im (red)  parts of eigenvalues as a function of momentum $k$ for  $\mu=\Delta$ (a) and $\mu=0.1\Delta$ (b). Shaded orange regions indicate zero real energies (BBFAs) with their borders marking the EPs.  Dotted gray, dashed-dot magenta, brown, and green curves show the absolute value of the overlaps  $|\bra{\psi_{e_1}}\ket{\psi_{e_{2}}}|$ and $|\bra{\psi_{e_\nu}}\ket{\psi_{h_\nu'}}|$, with $\nu,\nu'=1,2$, while cyan curve indicates   $\bra{\psi_{e_{\nu}(h_{\nu})}}\ket{\psi_{e_\nu(h_{\nu})}}=1$. Parameters as in Fig.\,\ref{Fig4}.   
	}
\label{Fig2Appendix} 
\end{figure}

It is thus clear that $\{E_{e_\nu},E_{h_\nu}\}$ and their respective eigenvectors $\{\Psi_{e_\nu},\Psi_{h_\nu}\}$ coalesce at $-k_{\rm EP_*}^{+}$. We illustrate this in Fig. \,\ref{Fig2Appendix}, where we replot Fig.~\ref{Fig4}(a,c) in the main text, now also with the eigenvector overlaps $\langle \Psi_{e_{1}}|\Psi_{e_{2}}\rangle$, $\langle \Psi_{e_{1}}|\Psi_{h_{2}}\rangle$, $\langle \Psi_{e_{1}}|\Psi_{h_{1}}\rangle$, and $\langle \Psi_{e_{2}}|\Psi_{h_{2}}\rangle$. Clearly, at the two-fold EPs, at the borders of the orange regions, the overlap $\langle \Psi_{e_{\nu}}|\Psi_{h_{\nu}}\rangle$ reaches one, indicating that $ \Psi_{e_{\nu}}$ and $  \Psi_{h_{\nu}}$ become parallel, see brown and green dashed curves. We verify that this is indeed the case by directly inspecting the eigenvector structure where, at the EPs, we obtain $ \Psi_{e_{\nu}}=\Psi_{h_{\nu}}$.   We also note that $  \Psi_{e_{1}}$ and $  \Psi_{e_{2}}$ become parallel at another EP at finite real energy, whose discussion has recently been reported in Ref.\,\cite{Cayao2023}. Therefore, we conclude that  $\{E_{e_\nu},E_{h_\nu}\}$, and their eigenvectors $\{\Psi_{e_\nu},\Psi_{h_\nu}\}$, coalesce at  the two-fold EPs discussed in Section \ref{section4}.

%%%%%%%%%%%%%%%%%%%%%%%%%%%%%%%
%                                  REFERENCES                                  %
%%%%%%%%%%%%%%%%%%%%%%%%%%%%%%%
\bibliography{biblio}

\end{document}